\documentclass[twocolumn,  
showpacs,  
preprintnumbers,  
aps,  
prd,  
a4paper,  
nofootinbib,  
tightenlines,  
floats, floatfix  
]{revtex4}

\usepackage{amsmath}
\usepackage{amsfonts}
\usepackage{graphicx,color}

\newcommand{\nn}{\nonumber \\}
\newcommand{\bea}{\begin{eqnarray}}
\newcommand{\ena}{\end{eqnarray}}

\newcommand{\LL}{{\cal L}}

\newcommand{\MM}{{\cal M}}

\renewcommand{\c}{\gamma}
\renewcommand{\d}{\delta}

\renewcommand{\O}{\Omega}

\renewcommand{\o}{\omega}

\begin{document}


\title{Nucleosynthesis constraint on Lorentz invariance violation in the neutrino sector}

\author{Zong-Kuan Guo}
\email{guozk@itp.ac.cn}
\affiliation{State Key Laboratory of Theoretical Physics, Institute of Theoretical Physics,
Chinese Academy of Sciences, P.O. Box 2735, Beijing 100190, China}

\author{Jian-Wei Hu}
\email{jwhu@itp.ac.cn}
\affiliation{State Key Laboratory of Theoretical Physics, Institute of Theoretical Physics,
Chinese Academy of Sciences, P.O. Box 2735, Beijing 100190, China}



\date{\today}

\begin{abstract}
We investigate the nucleosynthesis constraint on Lorentz
invariance violation in the neutrino sector which influences
the formation of light elements by altering the energy
density of the Universe and weak reaction rates prior to
and during the big-bang nucleosynthesis epoch.
We derive the weak reaction rates in the Lorentz-violating
extension of the standard model.
Using measurements of the primordial helium-4 and deuterium
abundances, we give a tighter constraint on the deformed
parameter than that derived from measurements of the cosmic
microwave background anisotropies.
\end{abstract}

\pacs{14.60.St, 98.80.Es }

\maketitle


Neutrino oscillation experiments have shown that there are small
but non-zero mass squared differences between three neutrino
mass eigenstates (see Ref.~\cite{gon07} and reference therein).
However, neutrino oscillations cannot provide absolute masses
for neutrinos. Cosmology provides a promising way
to constrain the total mass of neutrinos by the gravitational
effect of massive neutrinos on the expansion history near the
epoch of matter-radiation equality~\cite{kom10} and on the
formation of large-scale structures in the Universe~\cite{elg02}
(see also Ref.~\cite{won11} for a review).
Recently, a 3$\sigma$ detection of non-zero neutrino masses is
reported using new measurements of the cosmic microwave background
(CMB) anisotropies from the south pole telescope and Wilkinson
microwave anisotropy probe (WMAP), in combination with low-redshift
measurements of the Hubble constant, baryon acoustic oscillation
feature and Sunyaev-Zel'dovich selected galaxy clusters~\cite{hou12}.

These observations establish the existence of physics beyond
the standard model of particle physics. Another possible signal
of new physics is violation of Lorentz symmetry.
The possibilities of Lorentz invariance violation were
considered in string theory~\cite{kos89}, standard model
extension~\cite{coll98}, quantum gravity~\cite{ame02},
loop gravity~\cite{alf99}, non-commutative field theory~\cite{car01},
and doubly special relativity theory~\cite{mag01}.
Searches for Lorentz invariant violation with neutrinos have
been performed with a wide range of systems~\cite{kos11}.
Although present experiments
confirm Lorentz invariance to a good precision, it can be broken
in the early Universe when energies approach the Planck scale.
Cosmological observations provide a possibility to test such a
symmetry at high energies.

Recently, measurements of the CMB power spectrum were used to
probe Lorentz invariant violation in the neutrino
sector~\cite{guo12}. Lorentz invariant violation affects not
only the evolution of the cosmological background but also the
behavior of the neutrino perturbations.
The former alters the heights of the first and second peaks in
the CMB power spectrum, while the latter modifies the shape of
the CMB power spectrum. These two effects can be distinguished
from a change in the total mass of neutrinos or in the effective
number of neutrinos. The seven-year WMAP data in combination
with lower-redshif measurements of the expansion rate were used
to put constraints on the Lorentz-violating term.
However, the resulting constraints suffer from a strong correlation
between the Lorentz-violating term and the dark matter density
parameter~\cite{guo12}.

In this letter, we use current big-bang nucleosynthesis (BBN) data
to constrain Lorentz invariance violence in the neutrino sector.
There are two effects of Lorentz invariant violation on BBN.
The first is a correction to the weak reaction rate in the
Lorentz-violating standard model extension, which governs the
neutron-to-proton ratio at the onset of BBN.
The second is a change in the total energy density of the Universe.
Since the abundances of the light elements produced during BBN
depend on the competition between the expansion rate of the
Universe and the nuclear and weak reaction rates, the BBN
predictions depend on the Lorentz-violating term.
In particular the BBN-predicted abundance of helium-4 is very
sensitive to the deformation parameter.


We focus on Lorentz invariance violence only in the neutrino sector
and consider the following deformed dispersion relation
\bea
E^2 = m^2 + p^2 + \xi \, p^2,
\label{ddr}
\ena
where $E$ is the neutrino energy, $m$ the neutrino mass,
$p=(p^ip_i)^{1/2}$ the magnitude of the 3-momentum, and $\xi$
the deformation parameter characterizing the size of Lorentz
invariance violation. The dispersion relation implies that
there are departures from Lorentz invariance in the neutrino
sector if $\xi \neq 0$.
Such a deformed dispersion relation was constructed in the
framework of conventional quantum field theory~\cite{col98}
and derived in the Lorentz-violating extension of the
standard model~\cite{kos00}.

Here we have to point out that the dispersion relation for
neutrinos given in (\ref{ddr}) is not very general.
It neglects neutrino oscillations, possible species dependence,
anisotropies associated with violations of rotation symmetry,
and CPT violation. As shown recently by Kostelecky and Mewes,
all of these are possible~\cite{kos11}.
The model considered in this paper is one of many possible
Lorentz-violating theories.


The number density $n_\nu$ and energy density $\rho_\nu$ for
massive neutrinos with (\ref{ddr}) are given by~\cite{guo12}
\bea
\label{nd}
n_\nu &=& g_{\nu}\int \frac{d^3{\bf p}}{(2\pi)^3} f_{\nu}(E) \;, \\
\rho_\nu &=& g_{\nu}
 \int \frac{d^3{\bf p}}{(2\pi)^3} E f_{\nu}(E) \;,
\label{pd}
\ena
where $g_{\nu}=2$ is the number of spin degrees of freedom.
The phase space distribution for neutrinos is the Fermi-Dirac
distribution
\bea
f_{\nu}(E) = \left[1+\exp(E/T_{\nu})\right]^{-1},
\ena
where $T_{\nu}$ is the neutrino temperature.
Thus the number and energy density can be written as
$n_\nu=(1+\xi)^{-3/2}n^{(0)}_\nu$ and
$\rho_\nu=(1+\xi)^{-3/2}\rho^{(0)}_\nu$, where $n^{(0)}_\nu$
and $\rho^{(0)}_\nu$ are the standard number and energy
density, respectively.
Increasing $\xi$ decreases both the number and energy density.
The former leads to a reduced rate of the weak reaction prior
to and during the BBN epoch since the reaction rate is
proportional to the neutrino number density, while the latter
results in a reduced expansion rate of the Universe. Therefore,
Lorentz invariant violation affects the nucleosynthesis of
light elements.

\begin{figure}[b!]
\begin{center}
\includegraphics[width=80mm]{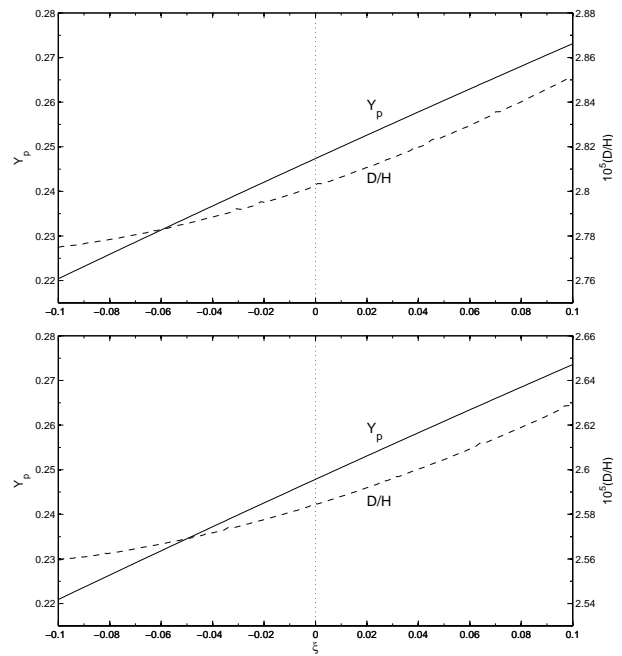}
\caption{The $^4$He mass fraction (solid curve) and D/H abundance
ratio (dashed curve) from BBN theory as a function of the
deformation parameter $\xi$ for $\O_b h^2=0.0213$ (the upper panel)
and $0.0224$ (the lower panel).
The vertical line corresponds to Lorentz invariance.}
\label{fig1}
\end{center}
\end{figure}

We turn now our attention to the details of the computation
of the weak reaction rate with Lorentz invariance violation
in the neutrino sector.
At early times when the temperature of the Universe was
$T \sim 100$ MeV, the number and energy density were dominated
by relativistic particles: electrons, positrons, neutrinos,
antineutrinos and photons. All of the particles were kept in
thermal equilibrium by the weak reactions
\bea
\label{enpe}
\nu_e+n  & \leftrightarrow & p+e^{-}, \\
e^{+} +n & \leftrightarrow & p+\bar{\nu}_e\;, \\
n        & \leftrightarrow & p+\bar{\nu}_e+e^{-}.
\label{npee}
\ena
When the expansion rate of the Universe exceeds the reaction
rate for $n \leftrightarrow p$ processes, the baryons become
uncoupled from the leptons. At the time the neutron-to-proton
ratio is frozen, which largely determines the primordial
helium mass fraction.
To estimate the neutron abundance at the onset of BBN
one has to compute the reaction rate.
As an example, let us consider the process $\nu_e+n \to p+e^{-}$.
The differential reaction rate per incident nucleon is
\bea
d\o &=& \sum_{\rm spins}\frac{|{\MM}|^2}{8 m_n m_p}
\frac{d^3 {\bf p}_{\nu}}{(2\pi)^3 2 E_{\nu}}f_{\nu}
\frac{d^3 {\bf p}_e}{(2\pi)^3 2 E_e}(1-f_e) \nn
&& 2\pi \d \left(E_{\nu} + m_n - m_p -E_e \right),
\label{drr}
\ena
where $|\MM|^2$ is the squared matrix element, to be summed
over initial and final state spins, $m_n$ and $m_p$ the neutron
and proton mass, respectively, ($E_e,{\bf p}_e$) the electron
four-momentum, and $f_e$ denotes the Fermi-Dirac statistical
distribution for electron. The process (\ref{enpe}) involves
the gauge boson $W$ as mediator. At tree level, one has
\bea
{\MM}=\frac{G_F}{\sqrt2}\,\bar{u}_p\c_{\mu}(C_V-C_A\c_5)u_n
\bar{u}_e (\c^{\mu}+c^{\mu\nu}\c_\nu)(1-\c_5)u_{\nu},
\ena
where $G_F$ is the Fermi coupling constant, $c_{\mu \nu}$
are the coefficients for Lorentz violation, and $C_V$, $C_A$
are the vector and axial coupling of the nucleon.
Here the coefficients $c_{\mu \nu}$ are defined to be traceless and isotropic.
After integrating (\ref{drr}) the reaction rate is
\bea
\label{rr}
\omega=\left[1-\frac38\,\xi
 -\frac{3\left(C_V^2-C_A^2\right)}{4\left(C_V^2+3C_A^2\right)}\,\xi\right]
 (1+\xi)^{-\frac32}\,\o^{(0)},
\ena
where $\o^{(0)}$ is the standard reaction rate per incident
nucleon derived in~\cite{esp98}.
The first factor on the right-hand side of~(\ref{rr})
arises from the neutrino propagator and the $e\nu W$ coupling in the Lorentz-violating
extension of the standard model~\cite{coll98} and the second
factor from the statistical distribution for neutrinos
(more general Lorentz-violating corrections involving electrons,
neutrinos, neutrons and protons were discussed in~\cite{lam05}).
At tree level, the differential reaction rates for the other
five processes in~(\ref{enpe})-(\ref{npee}) can be simply
derived from~(\ref{drr}) by properly changing the statistical
factors and the delta function determined by the energy
conservation for each reaction.
Therefore, the corrections to the conversion rate of neutron
into proton and its inverse are the same as in~(\ref{rr}).

From ~(\ref{rr}) we see that increasing $\xi$ reduces the
reaction rate, and thus the weak reactions freeze out at
earlier time, corresponding to a higher freeze-out temperature.
This leads to a larger neutron-to-baryon ratio at the onset
of BBN and thus a larger abundance of primordial $^4$He production.
On the other hand, increasing $\xi$ also reduces the
expansion rate of the Universe due to a decrease of the energy
density, which means the weak reactions freeze out at later
time without corrections to the reaction rate induced by the
deformed parameter. This therefore results in a lower helium-4
abundance. These two effects play opposite roles in the BBN
prediction for the helium-4 abundance.
The abundances of the other light nuclides weakly depend
on $\xi$ by changing the neutron-to-proton ratio and the
expansion rate.

Considering these corrections to both the reaction rate and
the expansion rate, we now estimate the freeze-out temperature,
$T_f$, determined by equating the expansion rate with
weak reaction rate.
In the Friedman-Robertson-Walker Universe, the expansion
rate obeys $H^2 = 8\pi G \rho/3$ where $\rho \propto T^4$
at early times. Thus, we have $H \propto (1-0.75\xi)T^2$.
Since the standard reaction rate in Eq.~(\ref{rr}) is roughly
given by $\o^{(0)} \propto T^5$~\cite{ber89}, we have
$\o \propto (1-1.80\xi) T^5$.
Setting $H \sim 4\o$ since the free-neutron decay process
and its inverse are neglected at the BBN epoch, one derives the
freeze-out temperature
\bea
T_f \sim (1+0.35 \xi) T^{(0)}_f,
\ena
where $T^{(0)}_f$ is the standard one. For a large $\xi$,
the weak reactions freeze out at a higher temperature.
This implies that effects caused by changing the reaction rate
dominate over those by changing the expansion rate due to
the Lorentz invariance violation in the neutrino sector.

In order to calculate the abundances of light elements
produced during BBN, we modified the publicly available
PArthENoPE code~\cite{pis07} to appropriately incorporate
the Lorentz-violating term in the neutrino sector.
Figure~\ref{fig1} shows the $^4$He mass fraction and D/H
abundance as a function of $\xi$ for $\O_b h^2=0.0213$ (the
upper panel) and $0.0224$ (the lower panel).
Both $Y_p$ and D/H increase as $\xi$ increases since the
effect of changing the reaction rate play a leading role.
Moreover, the dependence of $Y_p$ on $\xi$ is much larger,
relative to its observational uncertainties, than that of D/H.
Therefore, the primordial helium-4 abundance can provide a
sensitive probe of neutrino physics with Lorentz invariance
violation.


\begin{figure}[t!]
\begin{center}
\includegraphics[width=80mm]{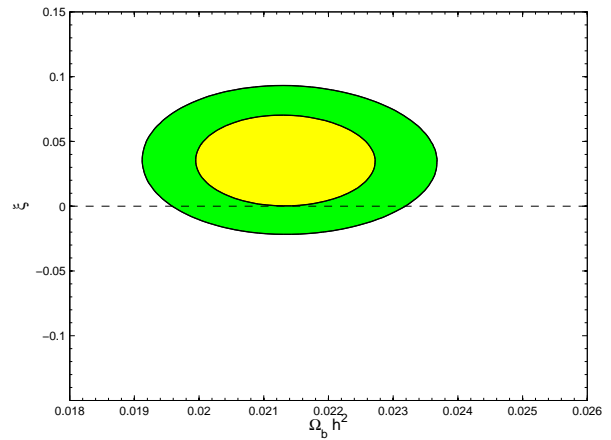}
\caption{Two-dimensional joint marginalized constraints (68\% and
95\% confidence level) on the deformation parameter $\xi$ and
physical baryon density $\O_b h^2$ from measurements of
$Y_p$ and D/H.
The dashed line corresponds to Lorentz invariance.}
\label{fig2}
\end{center}
\end{figure}

Assuming that there are three types of neutrinos with vanishing
chemical potentials in the Universe, the BBN-predicted primordial
abundances depend on only two parameters: $\O_b h^2$ and $\xi$.
As shown in Figure~\ref{fig1}, the abundance of deuterium is more
sensitive to the baryon energy density parameter but less sensitive
to the deformation parameter while that of helium-4 is more
sensitive to $\xi$ but less sensitive to $\O_b h^2$.
We use the observed primordial abundances of $^4$He and D in
combination to constrain these two parameters based on the
likelihood function
\bea
-2 \ln \LL = \frac{(Y_p-0.2565)^2}{0.006^2}
 + \frac{(\log[{\rm D/H}]+4.55)^2}{0.03^2}.
\ena
Here we adopt the estimate of the primordial helium mass fraction,
$Y_p=0.2565\pm0.0060$, derived in~\cite{izo10} using Monte Carlo
methods to solve simultaneously for many possible systematic effects,
based on 93 spectra of 86 low-metallicity extragalactic HII regions.
While some have employed a selected subset of these data for more
detailed analyses, the sources and magnitudes of systematic errors
have rarely been addressed. The measurement uncertainty in $Y_p$
is currently dominated by systematic errors.
For the primordial deuterium abundance, we use the value of
$\log[{\rm D/H}]=-4.55\pm0.03$ obtained in~\cite{pet08} from
measurements of the absorption lines of seven quasars at high
redshifts in low-metallicity hydrogen-rich clouds with low internal
velocity dispersions.
Besides $^4$He and $D$, $^3$He and $^7$Li are the other two
nuclides predicted in measurable quantities by BBN.
Since their post-BBN evolutions are complicated and their measurements
suffer from systematical uncertainties which are difficult to
quantify (for helium-3) or are poorly understood (for lithium),
the observed $^3$He and $^7$Li do not provide a reliable probe
of BBN at present, as discussed in~\cite{ste12}.
Thus, we do not include them in our constraints.

The $^4$He abundance is used to provide a constrain on the
deformation parameter while the D abundance is used to provide
a constrain on the baryon density parameter.
Using the combination of the $^4$He and D data, we find
$\xi=0.036\pm0.023$ and $\O_b h^2=0.0213\pm0.0009$ (68\% confidence level).
This estimated value of the deformation parameter is consistent
with Lorentz invariant $\xi=0$ within 95\% confidence level.
Compared to the results derived from the 7-year WMAP data
in combination with lower-redshift measurements of the expansion
rate~\cite{guo12}, BBN gives smaller uncertainties in $\xi$
by a factor of 4 because there is nearly no correlation
between the deformation parameter and the baryon density
parameter as shown in Figure~\ref{fig2}.
The estimate of $\O_b h^2$ is agreement with that from the
CMB data~\cite{kom10} with errors.


In summary, we have shown that the BBN puts strong constraint
on the deformed parameter in the Lorentz-violating extension
of the standard model, $\xi=0.036\pm0.023$.
Since the BBN-predicted abundance of helium-4 is very sensitive
to the deformed parameter but less sensitive to the baryon
energy density parameter, there is nearly no correlation
between the two parameters.
Our results indicate no significant preference for departure from
Lorentz symmetry in the neutrino sector in the early Universe.
Compared to previous constraints on the Lorentz-violating coefficient,
current BBN observation yields a weaker constraint.
As listed in Table XIII of~\cite{kos11}, the coefficient is
constrained down to $10^{-9}$ from time-of-flight measurements.
Cohen and Glashow have argued that the observation of neutrinos
with energies in excess of $100$ TeV and a baseline of at least
$500$ km allows us to deduce that the Lorentz-violating parameter
is less than about $10^{-11}$~\cite{coh11}.

\acknowledgments
We thank J.~Hamann, V.~A.~Kostelecky, O.~Pisanti and F.~Wang for
useful discussions.
Our numerical analysis was performed on the Lenovo DeepComp
7000 supercomputer in SCCAS.
This work is partially supported by the project of Knowledge
Innovation Program of Chinese Academy of Science,
NSFC under Grant No.11175225, and National Basic Research
Program of China under Grant No.2010CB832805.

\end{document}